\documentclass[twocolumn,nofootinbib]{revtex4-1}

\usepackage{graphicx}
\usepackage{caption}
\usepackage{amsmath,amssymb} 
\usepackage{latexsym}
\usepackage{mathrsfs}
\usepackage{bbold}
\usepackage{color}
\usepackage{slashed}
\definecolor{red}{rgb}{1,0,0}
\usepackage{cancel}

\definecolor{red}{rgb}{1,0,0}

\def\bea{\begin{eqnarray}} \def\eea{\end{eqnarray}}
\def\be{\begin{equation}} \def\ee{\end{equation}}

\newcommand{\Tr}{\text{Tr}}

\newcommand{\SO}{\text{SO}}
\newcommand{\SU}{\text{SU}}
\newcommand{\U}{\text{U}}

\newcommand{\promille}{%
  \relax\ifmmode\promillezeichen
        \else\leavevmode\(\mathsurround=0pt\promillezeichen\)\fi}
\newcommand{\promillezeichen}{%
  \kern-.05em%
  \raise.5ex\hbox{\the\scriptfont0 0}%
  \kern-.15em/\kern-.15em%
  \lower.25ex\hbox{\the\scriptfont0 00}}

\usepackage[colorlinks,linkcolor=blue,citecolor=blue,urlcolor=blue,linktocpage]{hyperref}
\newcommand{\hhref}[2][]{\href{http://arxiv.org/abs/#2#1}{arXiv:#2}}

\begin{document}
\title{Phenomenological aspects of supersymmetric composite Higgs models}
\author{Alberto Parolini}
\affiliation{SISSA and INFN, Via Bonomea 265, I-34136 Trieste, Italy}

\begin{abstract}
We study previously introduced models of pseudo Nambu-Goldstone boson Higgs from linearly realized minimal coset SO$(5)/$SO$(4)$ and matter in the fundamental in supersymmetric theories. Partial compositeness is at work for top and electroweak gauge fields. New states potentially relevant for LHC signatures are identified and we show how to reinterpret existing experimental results as exclusion bounds. The lightest colored particles, with a mass below the TeV, are fermionic and scalars top partners. We outline a viable mechanism originating masses of other Standard Model quarks: they result from the generation of dimension five operators in a non minimal flavor violating context. We study the impact of such operators on flavor processes and we show how experimental bounds are satisfied.
\end{abstract}

\maketitle

\setcounter{tocdepth}{1}
\tableofcontents

\section{Introduction}

After the discovery of a 126 GeV boson at the LHC \cite{Aad:2012tfa,Chatrchyan:2012ufa} the spectrum of the Standard Model (SM) has been entirely observed if this particle is identified with the Higgs boson. At the same time from an effective theory point of view such a value for the Higgs boson mass is highly unnatural and ameliorations of this problem are achieved if the SM is embedded in a larger theory. Anyhow explaining why no signal of such a theory has not revealed yet retaining the concept of naturalness is becoming more and more challenging as long as search for new physics phenomena does not yield positive results. After Run I of LHC a tension has grown between the need of Beyond the SM (BSM) physics close to the electroweak (EW) scale and bounds on BSM particles: the concept of naturalness has to be revised and we have to accept a certain level of Fine Tuning (FT). 

Among possible extensions of the SM Supersymmetry (SUSY) is the most studied one: SUSY partners, in particular of the Higgs boson and of the top quark,  should improve the ultraviolet (UV) behaviour of the theory. The bounds on the appearance of such partners are a classical example of the little hierarchy problem mentioned above. 

Another different approach is offered by theories in which the Higgs scalar is a bound state of a new strong dynamics \cite{Kaplan,Dugan:1984hq}: the UV divergencies responsible for the SM hierarchy problem would be cut off at a scale at which the Higgs is resolved in its constituents. In this way a minimal and unavoidable amount of tuning, namely the separation between the weak scale and this new scale typically of the order of the TeV, is accepted. A clever idea to dynamically generate this separation is to embed the Higgs boson in a strongly interacting theory, controlled by a large coupling $g_*$, in which it emerges as a pseudo Nambu Goldstone boson (pNGB) of a spontaneously broken global symmetry and hence as a tree level flat direction, and rely on radiative corrections to give it a potential capable to break the EW symmetry \cite{Agashe:2004rs}. 

To couple such a Higgs boson to SM fermions and gauge bosons partial compositeness \cite{Kaplan:1991dc,Contino:2006nn} is often advocated: SM states mix with bound states of this new sector and interactions with the Higgs are consequently induced. At the same time partial compositeness does not respect the global symmetry of the strong sector and therefore, together with any other source of explicit breaking, it is responsible for the pseudo nature of the NG boson: with the strong sector in isolation the Higgs in the effective action would remain a flat direction, being the global symmetry only spontaneously broken, without any explicitly violating term. Partial compositeness predicts the existence of partners with the same quantum number of SM fields. They live at the scale $g_*f$ but some of them have  to be lighter, as pointed out in effective models \cite{Matsedonskyi:2012ym,Redi:2012ha,Marzocca:2012zn,Pomarol:2012qf}, if we want to correctly reproduce the Higgs mass: notice that if this is not the case and the BSM sector is consistently characterized by a single mass scale the Higgs is generically expected to be too heavy. 

There has been a lot of research activity along this line, recently summarized in an up to date review \cite{Bellazzini:2014yua}.
Considerable results have been achieved employing holographic techniques: it is believed that models with additional compact space dimensions find their sensible UV completions in string constructions. Four dimensional purely fermionic completions, without unnatural scalars, have also been looked for \cite{Barnard:2013zea,Ferretti:2013kya,Ferretti:2014qta}, the main obstruction being the lack of techniques to tackle strongly coupled theories. In this paper we explore a different route and we exploit the virtues of supersymmetry, namely we consider infrared (IR) theories arising below a certain scale $\Lambda$ as Seiberg dual of some SQCD like theory. The SM gauge group is embedded in a larger flavor symmetry and SM gauge couplings are faint with respect to the gauge coupling of the SQCD, and in first approximation they can be neglected along the RG flow, with some exceptions. At sufficiently low energy, since the SQCD is in the so called magnetic free phase, SM gauge interactions are important. At very high scales, while the SQCD in the electric phase is asymptotically free the same is not true for the SM gauge group because of the presence of many new charged fields: we eventually have the appearance of Landau poles.

These models fall in a class of SUSY  Composite Higgs Models (CHM) studied in \cite{Marzocca:2013fza}. They consist of the following structure: SSM fields (without the Higgs and possibly without the right top), a second sector denoted as composite providing a NG boson with the quantum number of the SM Higgs and a third hidden sector communicating, without spoiling the spontaneous breaking of the global symmetry, a soft SUSY breaking needed for obvious experimental reasons. For the purposes of this discussion neither the origin nor the mediation of SUSY breaking ought to be further specified. Note that given the embedding of partial compositeness into the superpotential stops and left sbottoms are also realized as partially composite states.

Aim of this article is to investigate CHM in SUSY theories, in particular we focus on incarnations introduced in \cite{Caracciolo:2012je}. They are based on a minimal coset $\SO(5)/\SO(4)$ linearly realized with SM fields in the fundamental and top fermion fully or partially composite. For non top SM fermion masses we depict a different mechanism other than partial compositeness, driven also by the fact that their impact on the Higgs potential is negligible: we avoid a proliferation of partners that otherwise would dominate the running of the gauge couplings and make Landau poles appear at low energies, ruining the structure generating a pNGB Higgs. We instead assume the generation at some high scale of dimension five operators coupling a quark superfield bilinear with a pair of ``quarks" of the composite sector, flowing in the IR to the SM Yukawa couplings: a mass separation from the EW scale is favored. In generic CHM partial 
compositeness moderates flavor violations \cite{Csaki:2008zd}; in our configuration we do not introduce fermionic partners for quarks but on the other hand we inherit the flavor problem of SUSY, hence we assume a sufficient level of alignment for squark masses: in this situation we discuss how the presence of additional dimension five operators is harmless. We also elaborate on experimental signatures and consequent exclusions from collider experiments, namely from LHC data analysis. We show how to reinterpret current searches as bounds on supersymmetric CHM.

The paper is organized as follows: in section \ref{general_review} we recall the models introduced in \cite{Caracciolo:2012je}; the case of a partially composite right top is investigated in section \ref{top right elementare}, where the expected BSM particles' spectrum is presented: limits from available experimental observations are derived in section \ref{experimental pheno}. In section \ref{tadpole section} we discuss a deformation of the vacuum of the theory and we link it, in section \ref{flavor section}, to irrelevant operators responsible for non top masses. In section \ref{top right composto} we briefly show the numerical results for a case with fully composite right top field, explaining why the model suffers a severe tension. At the end of the paper, section \ref{Conclusions}, we draw our conclusions. In appendix \ref{sec:Spectrum} we report, for completeness, the whole spectrum of the main model analyzed in the text.

\section{General Structure of the Models}\label{general_review}

We briefly review the framework of \cite{Caracciolo:2012je}. Beyond the SM superfields there is a $\mathcal{N}=1$ SUSY gauge theory with $\SO(4)_m$ gauge group respecting a global symmetry $G_f=\SO(5)\times G$. It can be viewed as the low energy theory, via Seiberg duality, of a SQCD gauge theory based on $\SO(N)$ gauge group with $N_f=N$ electric quarks $Q_I$. A mass term
\be
W_{el}\supseteq m Q_aQ_a,\quad a=1,\ldots,5
\ee
leaves a $G_f\subseteq\SU(N_f)$ invariance. For $N\geq6$ the theory is asymptotically free and it flows to a IR free magnetic theory with $\SO(4)_m$ gauge group and a superpotential of the form
\be
W_{mag}\supseteq-\mu^2 M_{aa}+hq_IM_{IJ}q_J
\ee
where $q$ and $M$ are dubbed dual or magnetic quarks and mesons. The magnetic quarks $q_I$ are in the fundamental of the $\SO(4)_m$ gauge group, while the mesons $M_{IJ}$ are singlets. Both fields are composite in terms of the underlying 
degrees of freedom of the $\SO(N)$ UV theory. The coupling $h$ is not calculable within the duality recipe but a reasonable assumption is that it reaches a Landau pole at the same energy of the magnetic gauge coupling $g_{m}$. The parameter $\mu$ is defined as $\mu^2=-m\Lambda$ where $\Lambda$ is the dynamically generated scale and the hierarchy $m\ll\Lambda$ is required. 

The term $W_{mag}\supset-\mu^2 M_{aa}$ is responsible for a spontaneous breaking of both the global symmetry $G_f$ and of supersymmetry \cite{Intriligator:2006dd}.
Up to global rotations, the non-supersymmetric, metastable, vacuum is at
\be
\langle q_m^{n} \rangle  = \frac{\mu}{\sqrt{h}} \, \delta_m^n= \frac{f}{\sqrt{2}} \, \delta_m^n\,, 
\label{vuoto generale}
\ee
with all other fields vanishing. In eq.(\ref{vuoto generale}) we have decomposed the flavor index $a=(m,5)$, $m,n=1,\ldots,4$, and we have explicitly reported the $\SO(4)_m$ gauge index $n$ as well. 
SUSY is broken by the non vanishing F term of the meson $F_{M_{55}}=-\mu^2$. The vacuum (\ref{vuoto generale}) spontaneously breaks 
\be
\SO(4)_m\times \SO(5)\rightarrow \SO(4)_D\,,
\label{so9_6}
\ee
where $\SO(4)_D$ is the diagonal subgroup of $\SO(4)_m\times \SO(4)$. The global $G$ is left unbroken.
The six NGB's along the broken $\SO(4)_m\times \SO(4)$ directions, given by ${\rm Re}\, (q^m_n - q^n_m)$, are eaten by the 
$\SO(4)_m$ magnetic gauge fields $\rho_\mu$, that become massive, while the four NGB's along  $\SO(5)/\SO(4)_D$ remain massless
and  are identified with the four real components of the Higgs field.
At the linear level they are contained in Re $q_5^n$.

The SM vector fields are introduced by gauging a subgroup of the flavor symmetry group\footnote{The hypercharge is a combination of a $U(1)\subset\SO(5)$ and a $\U(1)_X\subset G$, where the $X$ charge is non vanishing for quarks.}
\be
G_f\supseteq \SU(3)_c\times \SU(2)_{0,L}\times \U(1)_{0,Y}\,.
\label{so9_8}
\ee
The $\SU(2)_{0,L}\times \U(1)_{0,Y}$ gauge fields introduced in this way are not yet the SM gauge fields, 
because the flavor-color locking given by the vev eq.(\ref{vuoto generale}) generates a mixing between the $\SO(4)_m \cong \SU(2)_{m,L} \times \SU(2)_{m,R}$ magnetic gauge fields and the elementary gauge fields.
The massless combination is identified with the actual SM vector fields.

The four uneaten NGB $h^{\hat{a}}$ can be collected within the matrix
\be\label{eq:Umatrix}
U=\exp\left(\frac{i\sqrt{2}}{f}h^{\hat{a}}T^{\hat{a}}\right)
\ee
where $T^{\hat{a}}$ are the four broken generators ($5\times5$ skew-symmetric hermitian matrices satisfying $\Tr\,T^{\hat{a}}T^{\hat{b}}=\delta^{\hat{a}\hat{b}}$, see appendix A of \cite{Caracciolo:2012je}) and $f$ is the decay constant of the $\sigma$-model.
The Higgs, being a NG boson of a spontaneous breaking of a global symmetry, could be removed from the non derivative part of the action with a field redefinition: however the symmetry is not exact and it is explicitly broken by the SM gauge group and by the coupling with the  top (super)field, leading to a potential for the Higgs field. We call gauge and matter contribution respectively the contributions to this potential proportional to powers of the gauge coupling and of the mixings respectively.
We follow the nomenclature and the notation of \cite{Marzocca:2013fza}, in particular we parametrize the Higgs potential in the unitary gauge as
\be
V(h)=-\gamma s_h^2+\beta s_h^4+\ldots\,.
\ee
where $s_h=\sin\frac{h}{f}$. We restrict to solutions with a minimum with fixed value $\xi=0.1$, defined as $\xi=\sin^2\frac{\langle h\rangle}{f}\simeq\frac{\gamma}{2\beta}$. Since we have $m_h^2\sim\xi\beta$ we can trade the pair $\{\xi,m_h\}$ for the pair $\{\gamma,\beta\}$.

\section{$\SO(11)$ Model with elementary $t_R$}\label{top right elementare}
\subsection{Structure of the Lagrangian}
An explicit realization of a SUSY model where $t_L$ and $t_R$ are elementary to start with is based on a $\SO(11)$ gauge theory with $N_f=11$ flavors, introduced in \cite{Caracciolo:2012je}. It is a model with vector resonances as described in \cite{Marzocca:2013fza} and, in their notation, it dynamically realizes the condition $h/\sqrt{2}=\lambda_R=\lambda_L$\footnote{These are not $\lambda_{L,R}$ appearing in the following equations. We instead kept $h$ with the same meaning.}. These couplings control the top mass and their strength is directly related to the appearance of some top partners: the weaker they are the lighter these top partners have to be to reproduce the top mass. In the present setup they are naturally stronger than in the general work of \cite{Marzocca:2013fza}. This observation will allow us, in the next subsection, to numerically explore a different and wider zone in parameter space than the one inspected in \cite{Marzocca:2013fza}.

The superpotential of the composite sector is
\begin{eqnarray}
W_{el} &=& m Q^a Q^a -\frac{\lambda_1}{2\Lambda_L} (Q^i Q^j)^2-\frac{\lambda_2}{2\Lambda_L} (Q^i Q^a)^2+\nonumber\\
&&+\lambda_L (\xi_L)^{ia} Q_i Q_a + \lambda_R (\xi_R)^{ia} Q_i Q_a\,.
\label{so11_1}
\end{eqnarray}
We split the flavor index $I$ ($I=1,\ldots, 11$) in two sets $I=(i,a)$, $a=1,\ldots,5$, $i=6,\ldots,11$. The last two terms in the superpotential 
are Yukawa couplings between the SSM fields $\xi_{L,R}$ and the fields $Q^I$ of the composite sector.
The fields $\xi_{L,R}$ encode $t_L$ and $t_R$ in covariant spurions.

When $\lambda_L=\lambda_R=0$,  the global symmetry is
\be
G_f = \SO(5) \times \SO(6) \,,
\label{so11_2}
\ee
with $\SO(5)$ and $\SO(6)$ acting on the $Q_a$ and $Q_i$ flavors, respectively.
The $\SO(11) $ theory becomes strongly coupled at the scale $\Lambda$. Below that scale, it admits a weakly coupled description
in terms of a magnetic dual $\SO(4)_m$ gauge theory with superpotential
\begin{eqnarray}\label{Wmagn}
W_{mag} &=& - \mu^2 M_{aa} -\frac12m_1 M_{ij}^2-\frac12 m_2  M_{ia}^2+\nonumber\\
&& + \epsilon_L (\xi_L)^{ia} M_{ia} + \epsilon_R (\xi_R)^{ia} M_{ia}+\nonumber\\
&&+h q_I M_{IJ} q_J\,,
\end{eqnarray}
where 
\begin{eqnarray}
\mu^2 &=& - m \Lambda\,, \ \ \ \ m_1 = \frac{\lambda_1 \Lambda^2}{\Lambda_L}\,, \ \ \ \ \ m_2 = \frac{\lambda_2 \Lambda^2}{\Lambda_L}\,, \nonumber\\ 
\epsilon_L &=& \lambda_L \Lambda\,, \ \ \ \ \ \epsilon_R= \lambda_R \Lambda\,,
\label{so11_4}
\end{eqnarray}
are the low energy parameters in terms of the microscopic ones. $\epsilon_L$ and $\epsilon_R$ are the couplings in front of the SUSY version of partial compositeness operators and $M_{ia}$ provide the needed superfield resonances. For simplicity, in the following we take all the parameters in eq.(\ref{so11_4}), including $h$, to be real and positive.

We add explicit soft SUSY breaking terms:
\begin{eqnarray}\label{SO(6) Vsoft}
V_{soft}&=&\widetilde m_{t_L}^2 |\widetilde q_L|^2 +\widetilde m_{t_R}^2 |\widetilde t_R|^2  + (\frac 12 M_\alpha  \lambda_\alpha \lambda_\alpha+h.c.)+\nonumber\\
&& + \widetilde m_1^2 |M_{ia}|^2 + \widetilde m_2^2 |M_{ab}|^2 + \widetilde m^2_3 |q_{i}|^2 + \nonumber\\
&&- \widetilde m^2_4 |q_{a}|^2 - \widetilde m^2_5 |M_{ij}|^2\,,
\end{eqnarray}
Because of these soft terms the vacuum eq.(\ref{vuoto generale}) gets modified to
\be\label{vev magn quarks}
\langle q_n^m\rangle=\frac{\tilde{\mu}}{\sqrt{h}}\delta_n^m= \frac{f}{\sqrt{2}} \, \delta_m^n,\quad\tilde{\mu}=\sqrt{\mu^2+\frac{\widetilde m_4^2}{2h}}\,.
\ee


\subsection{Numerical Analysis}

We show here the numerical results from an extensive scan in parameter space in the determination of the mass spectrum of the model, particularly the Higgs mass. We fix $\epsilon_R$ by requiring the correct top mass $m_t(1\text{ TeV}) \simeq 150 \text{ GeV}$ and then scan randomly for the other parameters searching for points with $\xi \simeq 0.1$. For any such point we then extract the Higgs mass from the exact potential and compute the full spectrum.

We find that the Higgs mass is distributed in the range $70 \text{ GeV} \lesssim m_H \lesssim 160\text{ GeV}$, peaking between $100-140\text{ GeV}$. The measured value $m_H \simeq 126\text{ GeV}$ is therefore a typical value for this model. For each point of the scan we obtain the FT computing numerically the logarithmic derivative of the logarithm of the Higgs mass with respect to all the parameters of the model, and taking the maximum value \cite{Barbieri:1987fn}. The FT ranges between $\sim 10$ and $\sim 300$, the typical value being around $50$, with no evident correlation with the value of the Higgs mass.

Let us now discuss some properties of the spectrum in the gauge sector and in the matter sector. The details of the particle content and analytic formulae can be found in Appendix \ref{sec:Spectrum}.

\subsubsection{Gauge Sector}\label{sect:spettro gauge}

The mass of the spin-1 resonances is given by $m_\rho = g_m f$, up to corrections of order $\mathcal{O}(g_{SM}/g_m)$ due to mixing with the elementary $\SU(2)_L \times \U(1)_Y$ gauge bosons. Considerations of metastability and perturbativity fix $g_m(f) \simeq 2.5$, which means $m_\rho \simeq 1950$ GeV for $\xi = 0.1$. Such values are still above the experimental limits from direct searches at the LHC \cite{CMS:2013vda,ATLAS:2013lma},\cite{Pappadopulo:2014qza} for limits not from experimental collaborations, but are in tensions with indirect bounds from the S parameter.

The lightest uncolored scalar resonance has a mass bounded from above by the same value as the vector resonance. It usually is the complex neutral singlet $M_{55}$ with a mass roughly around the TeV. With less frequency it is the $\text{Im}\,q_5^n$ or the lightest eigenstate of the symmetric part of $q_m^n$. 

Among the spin-1/2 states, the lightest one in our scan is usually a wino ($200 - 1200$ GeV) or the doublet $\tilde h_{u,d}$  arising from $q_5^n$ and $M_{5n}$ (around $1$ TeV) or the state in the $(1,3)$ of $\SU(2)_L\times \SU(2)_R$ coming from the magnetic gauginos $\rho$ and the fermions in the antisymmetric part of $q_m^n$ (in particular, the one $\tilde\rho^\pm_R$ with $Y = \pm 1$, $600 - 1300$ GeV, which does not mix with the elementary bino and the one $\tilde \rho_R^3$ which does mix, $200 - 1200$ GeV).

The goldstino, contained in  the superfield $M_{55}$, combines with the goldstino coming from the external SUSY breaking: a combination of the two will be eaten by the gravitino and the orthogonal will stay in the spectrum as a massive particle. The exact value for their masses depends on the F terms and we can have different mixed situations in collider experiments, leading to a cascades of decays from  neutralino to pseudogoldstino in turn decaying to the true goldstino, resulting in multiphoton events (and missing transverse energy) \cite{Ferretti:2013wya}.

\subsubsection{Matter Sector}
As in some non-SUSY CHM, the lightest colored fermion resonance is the exotic doublet with $Y=7/6$: the singlet with $Y=2/3$ coming from a mixture of the elementary $t_R$ and $M_{i5}$ is heavier, typically $\sim 1$ TeV, while the mass of the lightest fermion ranges up to $900$ GeV.

\begin{figure}[t]
\begin{center}
	\includegraphics[width=65mm]{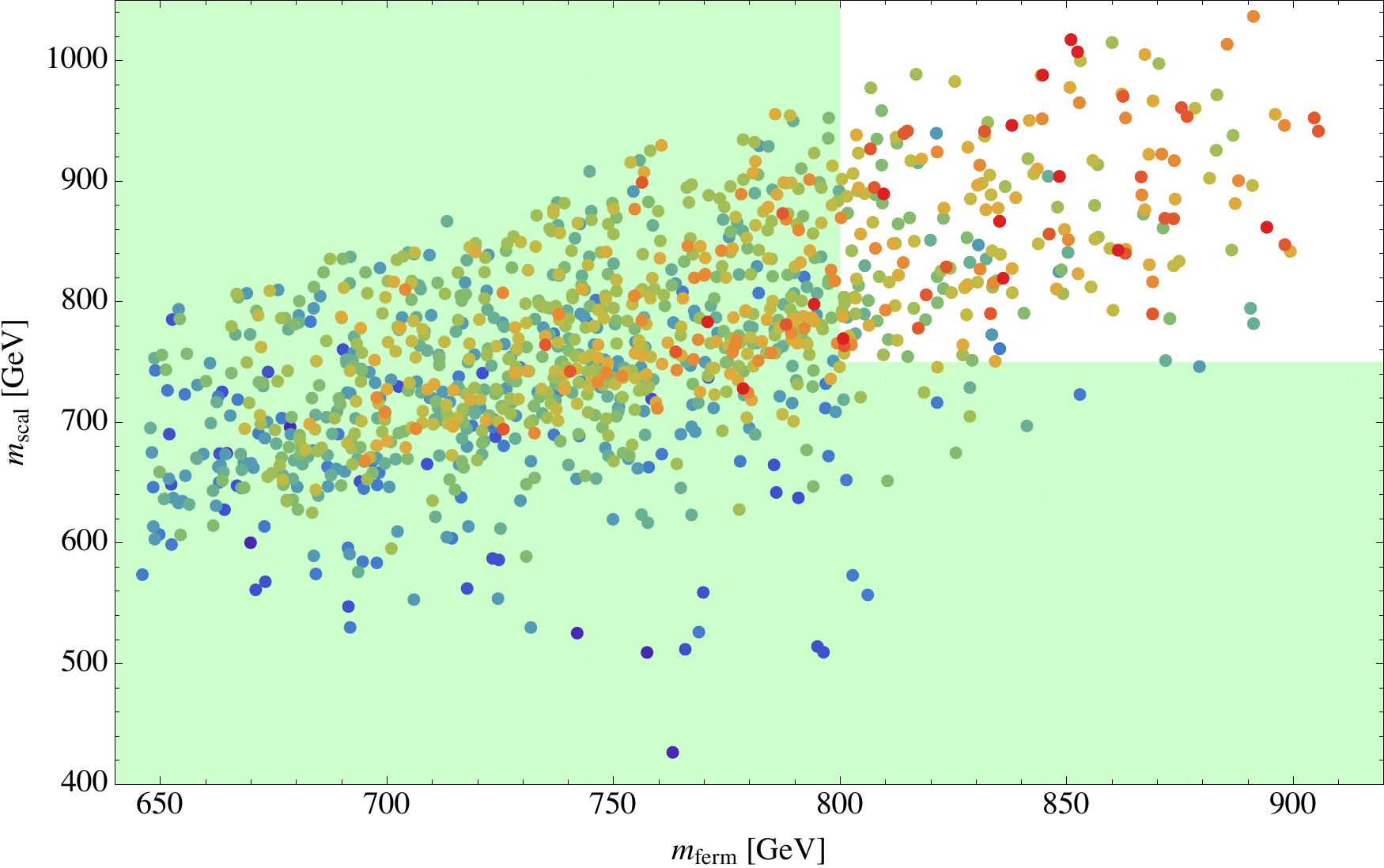}
	~~
	\includegraphics[width=7mm]{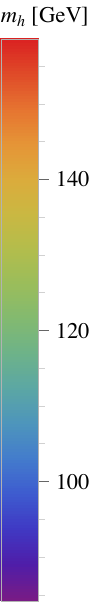}
\end{center}
\caption{Masses of the lightest colored fermions and scalar resonances. In shaded green we superimposed the exclusions discussed in section \ref{experimental pheno}. Colors represents Higgs mass as indicated aside.}\label{fig:colored_masses}
\end{figure}

In the case of colored spin-$0$ particles the spectrum contains stops and sbottoms as well as their composite partners and in absence of soft terms the discussion would proceed as for fermions. Taking into accout eq.(\ref{SO(6) Vsoft}) the lightest among the colored scalars is a resonance mixing either with $\tilde t_L$ or with $\tilde t_R$, respectively contained in a doublet with $Y=7/6$ or in a singlet with $Y=2/3$. Actually the whole bidoublet (a doublet with $Y=7/6$ and a doublet with $Y=1/6$) is almost degenerate in mass, the mass difference between the two doublets being $\lesssim 100$ GeV. The mass of the lightest scalar ranges from $600$ GeV to $1$ TeV. See fig. \ref{fig:colored_masses} for a scatter plot.

The gluino has a mass $M_3$ which does not enter the Higgs effective potential at one loop, therefore it can be heavier than the current bounds without affecting the FT of the model: contrary to what happens in the MSSM the EW scale is only logarithmically sensitive to stops masses \cite{Marzocca:2013fza}. It is also worth mentioning the existence of the chiral superfield $M_{ij}$ with supersymmetric mass $m_1$, singlet under the electroweak group and in the symmetric components of the $(3 + \bar{3})\times(3 + \bar{3})$ representation of $\SU(3)_c$. Since it does not enter the Higgs potential at one loop, its holomorphic mass $m_1$ is free in our setup and for consistency we take it $m_1<\Lambda\sim10$ TeV. In the following we will assume that it is heavy enough and neglect its phenomenology.

\subsubsection{A Benchmark Point}
We report here a benchmark point satisfying the bounds discussed in the following section selected from the set of the points collected with the numerical scan. We scanned over all the following parameters, allowing for randomly chosen $O(1)$ values, apart from $g_m$ which is determined by the ration $\mu/\Lambda=10^{-1}$ and $\epsilon_R$ which is fixed by the top mass. We have
\begin{eqnarray}
&&g_m(f)=2.5,\quad m_2=5.5\quad\epsilon_L=2.9,\quad\epsilon_R=3.3,\\\nonumber
&&\widetilde m_{t_L}=5.2,\quad\widetilde m_{t_R}=1.9,\quad M_1=2,\quad M_2=2.2,\nonumber\\
&&\widetilde m_{1}=0.5,\quad\widetilde m_{2}=0.9,\quad\widetilde m_{3}=0.3,\quad\nonumber\\
&&\widetilde m_{4}=0.6,\quad\widetilde{m}_{\lambda}=0.5\,,
\end{eqnarray}
where 
$m_\lambda$ is a soft Majorana mass for ${\SO(4)}_m$ gauginos and dimensionful parameters are expressed in units of $f=778$ GeV. The value of $h$ is fixed such that $0\simeq {g^{-1}_m(\Lambda)}={h^{-1}(\Lambda)}$. Taking into account QCD corrections at one loop we specify
\be
W=h qMq \supseteq 2h_m q_i M_{ia} q_a + h_g q_a M_{ab} q_b
\ee
since only superfields carrying an index $i$ are colored. For the given value of $m_2$ we have
\be
h_m(f)=1.9,\quad h_g(f)=1.4\,.
\ee
We then obtain $m_h=125.8$ GeV. The vector resonances have a mass of $2$ TeV, the lightest non colored scalar has a mass of $770$ GeV and it is the SM singlet we denoted with $M_{55}$, while other EW scalars are above the TeV and the lightest non colored fermion, besides the goldstinos, has a mass of $1.2$ TeV and the quantum numbers of a higgsino.

Among colored states the lightest fermion is the $Q=5/3$ exotic with a mass of $860$ GeV and the lightest scalar is a stop partner with a mass of $880$ GeV.

\section{Detection Bounds}\label{experimental pheno}

Given the features outlined in the previous section direct searches should concentrate on colored states, in particular fermions and scalars with exotic electric charge $5/3$. As stressed in Appendix \ref{sec:Spectrum} there is a consistent R-parity charges assignment (table \ref{table:R parity}): this implies, as usual, that scalar colored partners and EW fermion partners are pair produced and that the lightest among them is stable.

Fermionic top partners share the same R parity as elementary fields, because they mix with them; since they are a typical signature of CHM models \cite{DeSimone:2012fs} dedicated searches exist: $Q=5/3$ fermions are QCD pair produced and each of them decays to a $W$ boson and a top, in turn decaying to another $W$ and a bottom quark, therefore a good strategy is to look for events with two same sign leptons coming from the two $W$ bosons \cite{Contino:2008hi}. Since no excess has been observed CMS put a bound of $800$ GeV on the mass \cite{Chatrchyan:2013wfa} of these heavy fermions.

Turning to scalar particles the lightest is a stop partner and exotic scalars with $Q=5/3$ are typically a bit heavier. For such particles there are not dedicated searches; the main decay channels for them are wino plus the $Q=2/3$ top partner,  wino plus top (through its mixing with the heavy doublet) and gravitino plus fermionic $Q=5/3$ partner. The branching ratios depend on the details of the spectrum. In case of light charginos and heavy enough stop partners we can try to reinterpret the results for sbottoms pair produced and decaying into winos and tops: events with two b-jets and isolated same sign leptons are considered by CMS in \cite{Chatrchyan:2013fea} and a bound is set at $550$ GeV, well below the values found in the numerical scan.

In the model presented the scalar with $Q=5/3$ is almost degenerate with a full bidoublet of $\SO(4)$, namely with other scalars with $1/3$ and $2/3$ electric charge. As it happens for the fermionic partners Higgs vev insertions and the mixings $\epsilon_{L,R}\neq0$ affect the masses of these particles and remove this degeneracy inducing splitting of order $100$ GeV.
Thus we analyze limits on the masses of the other components of the bidoublet: in particular the scalar with $Q=2/3$ would behave similarly to a stop with decoupled gluinos.
Bounds on stops decaying into top and neutralino or bottom and chargino in events with one isolated lepton are derived by CMS from the full 19.5 fb$^{-1}$ dataset and stops are excluded with a mass approximatively below $650$ GeV \cite{Chatrchyan:2013xna}. 

Also CMS collaboration provides a stronger bound, of $750$ GeV \cite{CMS:2013cfa}, on pair produced stops each decaying into top and neutralinos using razor variables.

\begin{figure}[t]
\begin{center}
	\includegraphics[width=85mm]{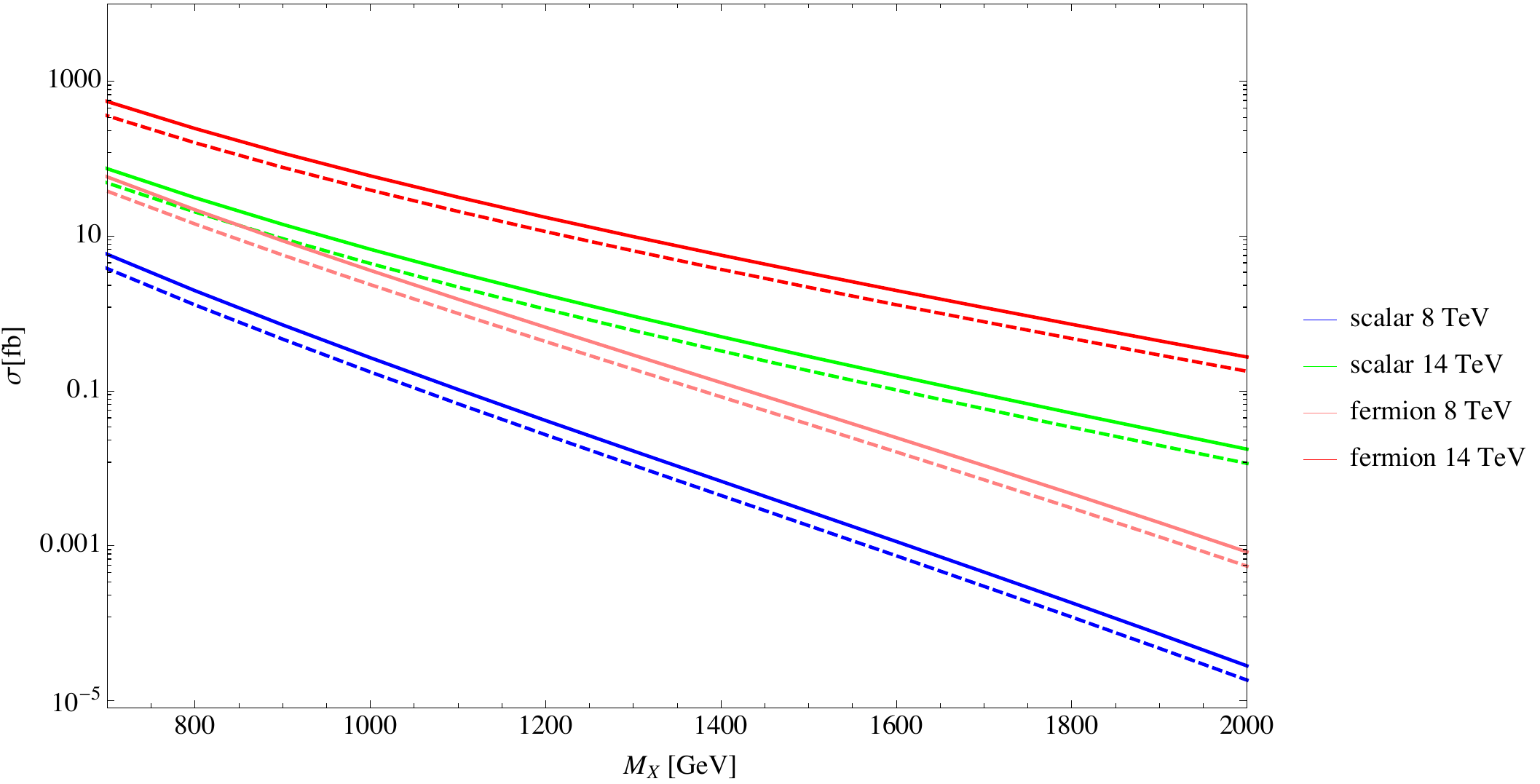}
	\end{center}
\caption{Pair production cross sections at LHC through QCD interactions. Dashed lines are the LO values, computed with M{\footnotesize AD}G{\footnotesize RAPH} 5 \cite{Alwall:2011uj}, using CTEQ6L PDFs and the model produced with the package F{\footnotesize EYN}R{\footnotesize ULES} \cite{Alloul:2013bka}; solid lines are the NLO, using a common $K_{NLO}=1.5$.}\label{fig:cross_section}
\end{figure}

Finally we stress that the simultaneous presence of fermions and scalars in the same mass range can strengthen the respective exclusion limits. Also in our setup multiple scalar stop partners appear (see Appendix \ref{sec:Spectrum}) and each of these can be produced and decay at the LHC thus heightening the number of expected events and consequently the exclusion bounds. 
We denote with $\sigma(M)$ the pair production cross section of one scalar top partner with mass $M$ and with $M_{excl,n}$ the excluded mass in case of $n$ identical scalars; if we assume a BR $=1$ in top and neutralino we estimate
\begin{eqnarray}
n\, \sigma(M_{excl,n})&=&\sigma(M_{excl,1})\quad\Rightarrow\nonumber\\M_{excl,n}&=&\sigma^{-1}(\frac{\sigma(M_{excl,1})}{n})
\end{eqnarray}
assuming that the production cross section for $n$ particles is just $n$ times the case with a single scalar in the spectrum: we neglect decay chains and mutual interactions which deserve a dedicated study. We numerically have
\be
\frac{M_{excl,n}-M_{excl,1}}{M_{excl,1}}\simeq0.1\quad\mbox{for}\quad n=2,3.
\ee

Turning to non colored states the lightest particles are fermions with quantum numbers of EW gauginos or higgsinos. As recently summarized in \cite{Flowerdew:2014Moriond} limits on charginos and neutralinos pair produced have been set by ATLAS \cite{Aad:2014nua} and CMS \cite{CMS:2013jfa}: with all sleptons and sneutrinos decoupled they set limits at 350 GeV from events with three or more leptons in the final state. This analysis also allows CMS to put bounds on sbottoms and excludes at $95\%$ CL masses below 570 GeV.

\begin{figure}[t]
	\includegraphics[width=85mm]{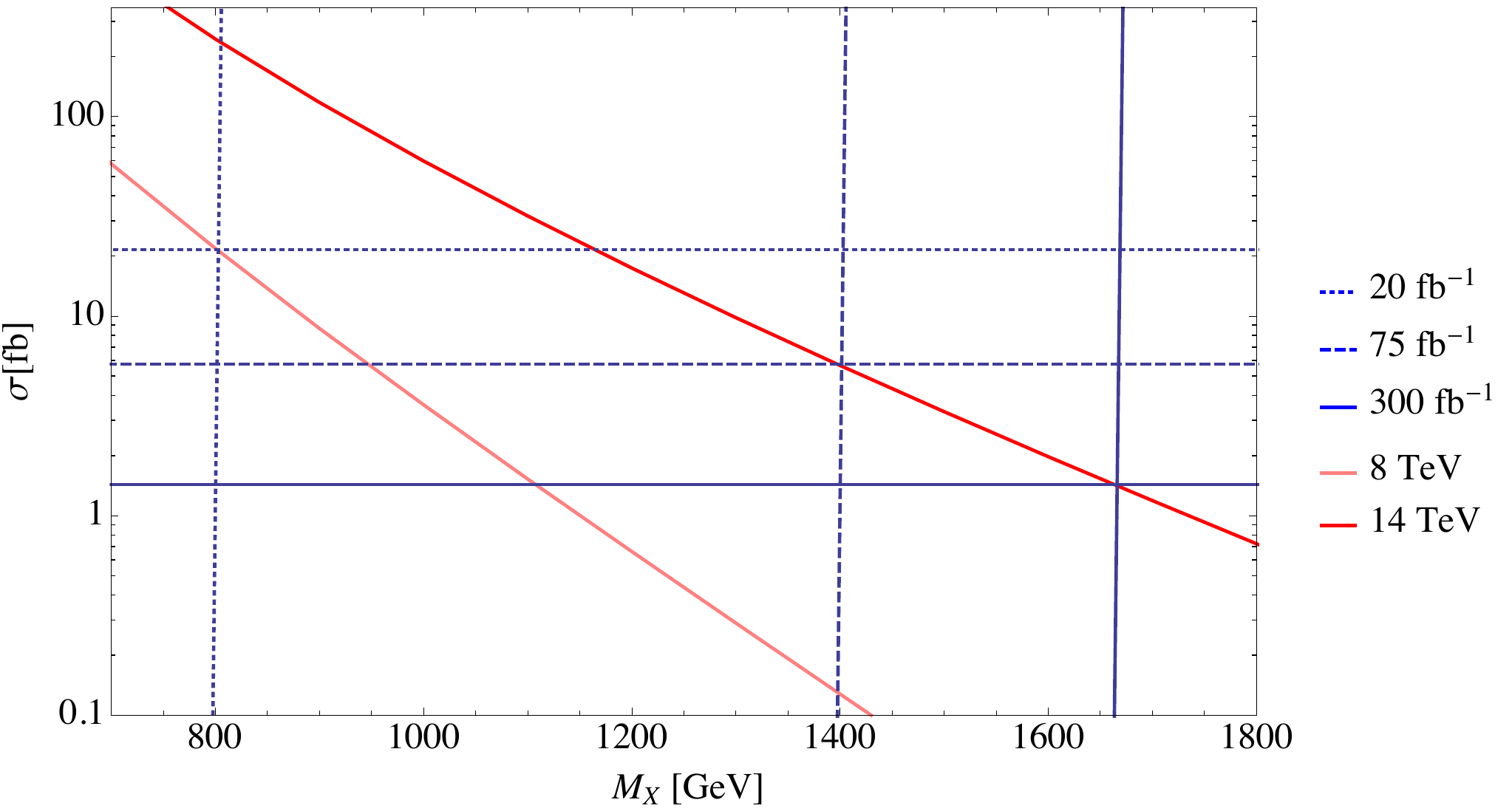}
	\includegraphics[width=85mm]{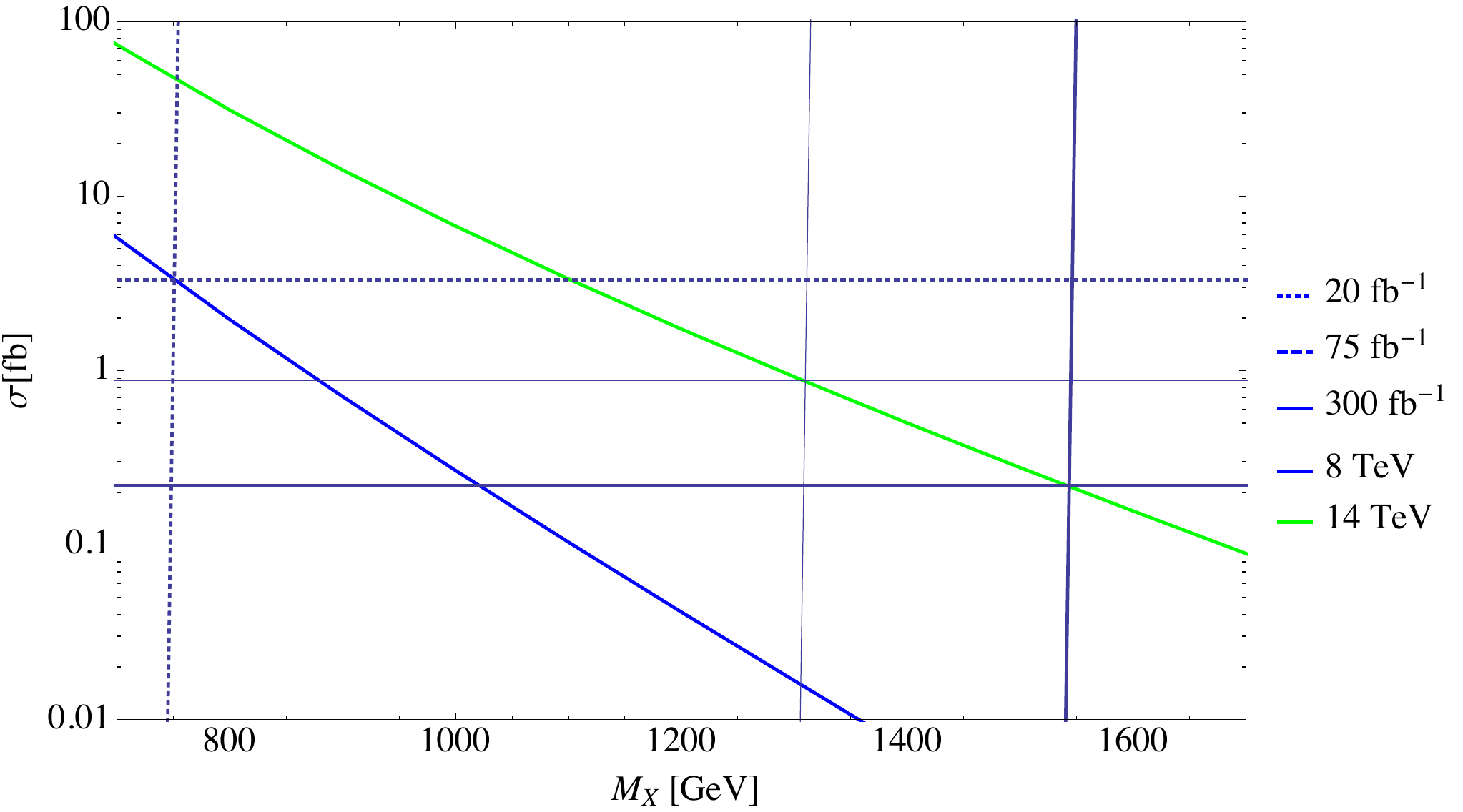}
\caption{Expected exclusion bounds on masses of the lightest fermionic (left panel) and scalar (right panel) top partners from $75$ (dashed line) and $300$ (solid line) fb$^{-1}$ at $\sqrt{s}=14$ TeV. The dotted lines correspond to present bounds at 800 and 750 GeV discussed in the text.}\label{fig:projections}
\end{figure}

Projections for exclusion limits for scalar and fermionic top partners from LHC at a center of mass energy of $14$ TeV can be obtained simply rescaling integrated luminosities\footnote{ATLAS published projections for future sensitivities in \cite{ATLAS:2013future}.}. Fig. \ref{fig:projections} clearly shows that higher luminosities, and higher center of mass energies, data will probe the relevant part of the parameter space. We expect they will be able to exclude exotic $5/3$ charge fermions up to 1400 (1650) GeV and scalars up to 1300 (1550) GeV with data corresponding to an integrated luminosity of $75$ ($300$) fb$^{-1}$, assuming BRs $=1$ for $Q=5/3$ fermions into $W$ and top; the $Q=2/3$ scalar is assumed to decay only to top and neutralino so to apply the analysis of \cite{CMS:2013cfa}. We also point out that for fermions the single production becomes more important than the pair production increasing the partner's mass and it takes over for heavy masses generically in the range $700$ - $1000$ TeV, both at $8$ and $14$ TeV. The single production does not go through QCD interaction and it is model dependent, it mainly happens as a $W$ and t fusion and it is generated by the mixing given by partial compositeness, therefore we do not expect the presence of SUSY to alter it significantly. We refer to \cite{Matsedonskyi:2014mna} for a more refined discussion.  

We conclude this section noting that we can interpret already existing experimental searches to exclude portions of the parameters space of the model in section \ref{top right elementare}: we expect future experiments, LHC at $14$ TeV will play a preponderant role, to further probe it and constrain it to regions with  higher level of FT.

\section{Soft Deformation of the Vacuum}\label{tadpole section}
In this section we turn back to the model building and we study soft deformations including, besides scalar and gaugino masses as in eq.(\ref{SO(6) Vsoft}), A and B terms for the couplings in the superpotential eq.(\ref{Wmagn}). They cannot be derived from the parameters of the electric theory, namely the techniques employed to follow the soft masses \cite{ArkaniHamed:1998wc} cannot be used in the case of A and B terms because the former can be computed exactly only in absence of any superpotential and thus they are expected to be valid up to perturbative corrections in couplings, while the latter are identically zero if the superpotential vanishes.

The most general soft terms may lead to unwanted tachyonic directions and we restrict to safe cases where they only modify the spectrum: this happens if they are not too large with respect to the holomorphic and soft masses. The presence of B terms for the electric quarks induces a term of the form
\be\label{eq:tadpole}
\mathcal{L}_{soft}\supseteq - {\mu^2} B_{\mu^2} \Tr\,M
\ee
which introduces a new qualitative feature, even for arbitrary small $B_{\mu^2}$. In fact the scalar potential now includes 
\begin{eqnarray}
V&\supseteq&{|2h M_{nn}q_m^n|}^2+{|hq_m^nq_m^n-\mu^2|}^2+\\
&&+(\mu^2B_{\mu^2}M_{aa}+h.c.)+\widetilde{m}_2^2{|M_{nn}|}^2-\widetilde{m}_4^2{|q_m^n|}^2\,.\nonumber
\end{eqnarray}
The vev of the magnetic quarks, eq.(\ref{vev magn quarks}), becomes
\be
\langle q_n^m\rangle=\frac{\tilde{\mu}}{\sqrt{h}}\delta_n^m\,,
\ee
where
\be
\tilde{\mu}\simeq\sqrt{\mu^2+\frac{\widetilde m_4^2}{2h}}-\frac{h \mu^4 (B_{\mu^2})^2}{\sqrt{\mu^2+\frac{\widetilde m_4^2}{2h}}{(4h\mu^2+2\widetilde m_4^2+\widetilde m_2^2)}^2}
\ee
expanding for small $B_{\mu^2}$: the true value for $\tilde{\mu}$ is a solution of a cubic equation. At the same time magnetic mesons also acquire a vev
\be\label{vev meson}
\langle M_{55}\rangle=X_5, \quad\quad\langle M_{mn}\rangle=X \delta_{mn}
\ee
where
\be
\left\{
\begin{array}{lll}
X_5&=&-\frac{\mu^2B_{\mu^2}}{\widetilde m_2^2}\\
X&=&-\frac{\mu^2 B_{\mu^2}}{4h\tilde{\mu}^2+\widetilde m_2^2}
\end{array}
\right.\,.
\ee
The spectrum of the theory gets modified but the only qualitative difference is about the Goldstone bosons: the four uneaten ones, identified with the Higgs, are now contained in the massless combination
\begin{equation}
\quad\cos\alpha \mbox{ Re }q_5^n+\sin\alpha\mbox{ Re }M_{5n}\,,
\end{equation}
where
\be
\sin\alpha=\frac{2(X-X_{5})}{f}\, .
\ee
Their kinetic term comes from ${|D_\mu q_a|}^2$ and ${|D_\mu M_{ab}|}^2$; at the non linear level they are described by a $\sigma$-model through a matrix $U$ as in eq.(\ref{eq:Umatrix}) with a new decay constant
\begin{equation}
f=\sqrt{\frac{2}{h}\left(\tilde{\mu}^2+2h{(X-X_5)}^2\right)}\, .
\end{equation}
In the limit of vanishing B term we have $X=X_{5}=0$ and $\cos\alpha=1$. For numerical analysis we work in the regime of small $B_{\mu^2}$, in particular we neglect its effects on the Higgs mass: this is consistent as long as the suppression between $B_{\mu^2}$ and other masses, both holomorphic and soft, is at least of the order of one loop effects.

In the next section we couple SM fermions pairs to the Higgs field exactly through its small component along the meson $M_{5n}$.

\section{Quark Masses}\label{flavor section}
\subsection{Generation}\label{subsec:quark masses operators}
The plain generalization of partial compositeness in our SUSY setup to all quarks and leptons does not work: it requires the presence of a large number of (super)partners, leading to tremendously large flavor symmetry of the composite sector and aggravating the problem of SM Landau poles. A sizeable contribution to the QCD beta function comes from the dual mesons in the adjoint of the flavor symmetry, which contains as a subgroup ${\SU(3)}_c$. For $N$ partially composite quarks we have a global ${\SO(6)}^N$, included in a simple $\SO(6N)$ global symmetry of the strongly interacting SQCD: the presence of such a large group and the presence of mesons in its adjoint representation make impracticable the extension of partial compositeness even only for the bottom. Therefore we abandon it for all the fermions but the top. Ordinary non SUSY models are less touched by this problem and they can accommodate partners for all quarks without introducing Landau poles below $4\pi f\simeq\Lambda$, the UV cutoff. Then they can be completed, at least in principle, by models with a non simple group of global symmetries, in contrast to what happens in our setup.

We need a further explicit breaking of the $\SO(5)$ global symmetry, proportional to two matrices $\lambda_U^{AB}$ and $\lambda_D^{AB}$ where $A,B=1,2,3$ are family indices. The extension to leptons through another pair of matrices is straightforward. Deformations in the electric superpotential\footnote{The simplest way to induce these operators, as well as the ones in eq.(\ref{so11_1}), is through the exchange of heavy chiral superfields, schematically $W=\lambda^{AB}\xi^A\Phi\xi^B+Q\Phi Q+\frac{\Lambda_L}{2}\Phi^2$.}
\begin{eqnarray}\label{Wel quark masses}
W_{el}&\supseteq&\frac{\lambda_U^{AB}}{\Lambda_L}{\left(\xi_{L,U}^{ia}\right)}^A{\left(\xi_U^{ib}\right)}^BQ_aQ_b+\\&&+\frac{\lambda_D^{AB}}{\Lambda_L}{\left(\xi_{L,D}^{ia}\right)}^A{\left(\xi_D^{ib}\right)}^BQ_aQ_b\nonumber
\end{eqnarray}
generate Yukawa terms if the dual mesons $M_{ab}\sim \frac{ Q_aQ_b}{\Lambda}$ get a vev, eq.(\ref{vev meson}). $\xi_{L,U}$ and $\xi_U $ are the spurionic embeddings of up type quarks in a fundamental of $\SO(5)$. $\xi_{L,D}$ and $\xi_D$ are the spurions for down type quarks and can be defined in analogy to the up case but with a different $X$ charge assignment, $X=-1/3$. 
The most general low-energy Lagrangian will contain
\be\label{eq:Lagra_masses}
\mathcal{L}\supseteq\bar{q}_L^A\varepsilon u_R^B H^c +\bar{q}_L^A \lambda_u^{AB} u_R^B H^c+\bar{q_L}^A\lambda_d^{AB}d_R^B H+...+h.c.
\ee
where $\lambda_{u,d}=\lambda_{U,D}^\dagger\frac{\Lambda}{\Lambda_L}\sin\alpha$\footnote{The dagger is only for notational convenience.} and the dots stand for higher dimensional operators: eq.(\ref{eq:Lagra_masses}) arises from the expansion in powers of $f^{-1}$ of $M_{ab}=U_{ca}\langle M_{cd}\rangle U_{db}$ once we make explicit the Higgs dependence through the matrix U defined in eq.(\ref{eq:Umatrix}). $H$ is the Higgs doublet $H={\left(H^{(+)},H^{(0)}\right)}^t=\frac{1}{\sqrt{2}}{\left(i h^1+h^2,-ih^3+h^4\right)}^t$.
Without loss of generality we can go to the top basis in which $\varepsilon\sim\frac{\epsilon_L^{A}\epsilon_R^{*B}}{f^2}$ is different from zero only for $A=B=3$: this is the term generated by the mixings in eq.(\ref{so11_4}). The second and the third terms in eq.(\ref{eq:Lagra_masses}) are the new operators responsible for the other quark masses. They have similarities with technicolor theories, where quark bilinears are coupled to an operator $H$ arising from a strongly interacting theory and responsible for EW breaking. The main differences are first that in our case the Higgs is protected by a shift symmetry and second that this coupling is not the dominant source for the top mass.

To estimate the size of these masses we restrict for a moment on a single generation:
\begin{equation}
\mathcal{L}\sim\lambda_D\sin\alpha\frac{\Lambda}{\Lambda_L}v\,\bar{b}_Rb_L+h.c.
\end{equation}
where $v=246$ GeV is the Higgs vev and $\Lambda$ is the crossover scale dynamically generated mentioned in section \ref{general_review}. If $\Lambda_L\sim10\Lambda$ the correct value for the bottom mass can be reached with $\lambda_D= O(1)$ and $\sin\alpha\sim0.1$. As discussed in \cite{Caracciolo:2012je} a natural value for $\Lambda$ is around $10$ TeV while $\Lambda_L$ can be chosen to be the scale of Landau poles for $\SU(3)_c$, in the region $10^2$ - $10^3$ TeV. Other quarks require smaller couplings: we do not explain neither the hierarchy among SM masses nor the hierarchical structure of the CKM matrix, we assume them and we only distinguish between the top, partially composite, and other quarks, elementary. This different origin of the masses results in the splitting between the top and other quarks, making them naturally live at two different scales. 

Finally we define $Y_d=\lambda_d$ and $Y_u=\lambda_u+\varepsilon$; they can be brought to diagonal form with
\be
Y_{u,d}\rightarrow Y_{u,d}^{diag}=V_{u,d}^\dagger Y_{u,d} U_{u,d}\,.
\ee
If we perform the transformations $V_d$, $U_u$ and $U_d$  we go to the basis in which $Y_d$ is diagonal, while $Y_u\rightarrow V_d^\dagger V_uY_u^{diag}$ and we can define the CKM matrix $V_{CKM}=V_u^\dagger V_d$. Thus
\be\label{lambda_u nMFV}
\lambda_u\rightarrow V_d^\dagger\lambda_u U_u=V_{CKM}^\dagger Y_u^{diag}-V_d^\dagger\varepsilon U_u\,.
\ee
Since the matrix $V_d^\dagger\varepsilon U_u$ has arbitrary entries the second term signals a departure from MFV. In the next subsection we elaborate on it and on its consequences.

\subsection{Flavor Constraints}
A number of processes involving transitions in flavor space, $\Delta F=0,1,2$, results in flavor and CP observables and they very often receive sizeable contributions from the presence of new physics, which in a broad class of CHM are mainly induced by the mixing of quarks with their partners. Despite the fact that these mixings are related to SM Yukawas the resulting suppression might be not enough for a generic composite sector and additional flavor symmetries are frequently postulated: a recent review is provided in \cite{Buttazzo:2014bka}. If light quarks are not partially composite, that is there are no mixings with bound states, these contributions are absent: this is almost the case for the interactions introduced in subsection \ref{subsec:quark masses operators}, we will formulate a more precise statement later in this subsection.

On the other hand our model exhibits the same tensions of the MSSM, due to the presence of sparticles around the TeV scale: squark mass matrices cannot be completely anarchic. Solutions to regulate the contributions to flavor processes are either to assume a certain level of degeneracy or alignment among squarks masses or to rely on some hierarchy between the first two generations and the third one, without threatening the naturalness, ending up with a scenario close to effective SUSY, depicted for instance in \cite{Brust:2011tb} where a discussion on flavor processes is also present. Correlations among new physics contributions in different processes could help in the future to distinguish among these possibilities \cite{Giudice:2008uk}. We derive our results with aligned squarks masses\footnote{Alignment is nicely realized in a variety of SUSY breaking mediation schemes: for instance in gauge mediation A-terms vanishes at the mediation scale.} and we allow for small misalignment treated in the mass insertion approximation.

We do not perform a full analysis of all existent bounds; we instead concentrate in what follows on the effects of the physics leading to the superpotential in eq.(\ref{Wel quark masses}): at the scale $\Lambda_L$ other operators are plausibly generated. Under the spurionic flavor group $\U(3)_q\times\U(3)_u\times\U(3)_d$ we assign the quantum numbers:
\begin{eqnarray}
q_L&\sim&(3,1,1),\quad u_R^c\sim(1,\bar{3},1),\quad d_R^c\sim(1,1,\bar{3}),\nonumber\\ \lambda_u&\sim&(3,\bar{3},1),\quad\lambda_d\sim(3,1,\bar{3})\, .
\end{eqnarray}
Compatibly with these charges, with gauge invariance and with the holomorphy of the superpotential we can write the following dimension five operators\footnote{At high energies the flavor spurions are $\lambda_{U,D}$ and not $\lambda_{u,d}$, the difference being a factor $\frac{\Lambda_L}{\Lambda\sin\alpha}\sim100$.}\textsuperscript{,}\footnote{$t^{\hat{A}}$ are the $\SU(3)_c$ generators such that $t^{\hat{A}}_{ij}t^{\hat{A}}_{kl}=\frac{1}{2}(\delta_{il}\delta_{jk}-\frac{1}{3}\delta_{ij}\delta_{kl})$.}
\begin{eqnarray}\label{eq:dangerous dim five}
W&\supseteq&\frac{a_1}{\Lambda_L}\left(u^c_R\lambda_Uq_L\right)\left(d^c_R\lambda_Dq_L\right)+\\&&+\frac{a_2}{\Lambda_L}\left(u^c_R\lambda_Ut^{\hat{A}}q_L\right)\left(d^c_R\lambda_Dt^{\hat{A}}q_L\right)=\nonumber\\
&=&\frac{Y^{ABCD}}{\Lambda_L}\left[a_1\left(u^c_{R,A}q_{L,B}\right)\left(d^c_{R,C}q_{L,D}\right)\right.\nonumber\\&&+\left.a_2\left(u^c_{R,A}t^{\hat{A}}q_{L,B}\right)\left(d^c_{R,C}t^{\hat{A}}q_{L,D}\right)\right]\nonumber
\end{eqnarray}
where $Y^{ABCD}=\lambda_U^{AB}\lambda_D^{CD}$. Dimension five operators in the MSSM are discussed in \cite{Pospelov:2005ks,Pospelov:2006jm,Antoniadis:2008es}: they results in, among other terms, contact interactions between two quarks and two squarks. We assign the couplings $\lambda_{u,d}$ to the vertices quark-squark-higgsino, neglecting deviations for tops, stops and left sbottoms. With higgsino exchange we draw one loop diagrams contributing to four fermions interactions, experimentally constrained by $\Delta F=2$ transitions in mesons. The resulting operator is
\begin{eqnarray}\label{eq:operatore dimensione sei}
&&
({\bar d}_{R,Ck} d_{L,Dl})({\bar d}_{L,Ei} d_{R,Fj})\, \frac{\lambda_u^{EA}\lambda_d^{BF}}{{(4\pi)}^2\widetilde m \Lambda_L}
\cdot\\
&&\cdot\left\{\delta_{ij}\delta_{kl}\left[Y^{ABCD}(a_1-\frac{a_2}{6})-Y^{ADCB}\frac{a_2}{2}\right]-
\scriptscriptstyle{\left(\begin{array}{c}
B\leftrightarrow D \\
j\leftrightarrow l
\end{array}\right)}
\right\}
\,,\nonumber
\end{eqnarray}
where $\widetilde m$ is a common soft mass for the squarks and the higgsino in the loop. For operators of the form
\be
\frac{c}{\Lambda_F^2}({\bar d}_{R} d_{L})({\bar d}_{L} d_{R})
\ee
the most stringent bounds come from kaons (the strongest is on the CP violating part). The Wilson coefficient computed from eq.(\ref{eq:operatore dimensione sei}) identically vanishes, even in the non MFV limit of eq.(\ref{lambda_u nMFV}). Non aligned squark masses at $\varepsilon=0$ results, in the mass insertion approximation, in (the hadronic matrix element with $j\leftrightarrow l$ is less significant by a factor of $3$ \cite{Ciuchini:1998ix,Bona:2007vi})
\begin{eqnarray}
\frac{c}{\Lambda_F^2}&\simeq&\frac{1}{{(4\pi)}^2\widetilde m \Lambda_L}A^2\lambda^5 {\left(\frac{\Lambda_L}{\Lambda\sin\alpha}\right)}^4y_d y_s y_t^2\delta\nonumber\\&\Rightarrow&
\left\{
\begin{array}{c}
\Lambda_F=1\mbox{ TeV}\\
c\simeq10^{-8}\delta\,\left(\frac{1 \mbox{\footnotesize TeV}}{\widetilde m}\right)\left(\frac{100\mbox{\footnotesize TeV}}{\Lambda_L}\right) 
\end{array}\right.
\end{eqnarray}
where for concreteness we fix $a_1=a_2=1$; $A$ and $\lambda$ are the parameters appearing in the Wolfenstein parametrization of the CKM matrix and $\delta$ measures the relevant misalignment of squarks: it is the mixing of the first two families left handed squarks normalized with a common mass ${\widetilde m}^2$, $\delta=\frac{{\left({\widetilde m}_{Q}^2\right)}_{1,2}}{{\widetilde m}^2}$.
From \cite{Isidori:2010kg} we easily read:
\be\label{eq:cboundK}
\mbox{Re}\,c<6.9\times10^{-9}\,,\quad\mbox{Im}\,c<2.6\times10^{-11} \quad\mbox{if}\quad\Lambda_F=1\mbox{ TeV}\,.
\ee
Bounds from box diagrams for different processes, with squarks and gluinos at $1$ TeV, are stronger, they set for $\delta$  an upper bound around $10^{-2}$, see \cite{Carrasco:2014uya} and references therein\footnote{Notice that in box diagrams down squarks run into the loop while loops with vertices from eq.(\ref{eq:dangerous dim five}) are sensitive to up squark mass insertions, constrained by box diagrams for $D-\bar{D}$ oscillation.}, resulting in $c\simeq10^{-10}$, below the bound eq.(\ref{eq:cboundK}) (the bound on the CP violating effect computed here is not fully satisfied, it needs $a_1\simeq a_2=O(10^{-1})$ or so, or smaller $\delta$). The numerical value of $c$ is of the same order also with general $\varepsilon$.

Similarly the up-type quarks are involved in D mesons oscillations: in this case the calculation is performed in the up-type mass basis, that is
\be
\lambda_u=Y_u^{diag}-V_u^\dagger\varepsilon U_u\,,\quad\lambda_d=V_{CKM}Y_d^{diag}\,.
\ee
The relevant operator has the same form as in eq.(\ref{eq:operatore dimensione sei}) with the exchange $u\leftrightarrow d$. In this case the coefficient is identically zero only if $\varepsilon=0$, and for $\varepsilon=O(1)$ it is controlled by ${\left(\frac{\Lambda_L}{\Lambda\sin\alpha}\right)}^2y_b^2A\lambda^2$; its value can be recast as
\be
\Lambda_F=1\mbox{ TeV},\quad c\simeq10^{-10}\,\left(\frac{1\,\mbox{\footnotesize TeV}}{\widetilde m}\right)\left(\frac{100\,\mbox{\footnotesize TeV}}{\Lambda_L}\right) 
\ee
with $a_1=a_2=1$, below the experimental constraints, $c<10^{-8}$ \cite{Isidori:2010kg}. Small squarks mass insertions do not change this numerical value\footnote{Bounds on down-type squark mass mixings are reported in \cite{Kersten:2012ed,Mescia:2012fg}.}.

Hence we can infer that the inclusion of eq.(\ref{eq:dangerous dim five}) does not reintroduce violations and does not hack the solution settled to avoid flavor problems.

At the same time a completely generic structure for $\lambda_{u,d}$ is disfavored: in fact although there are heavy fermionic partners only for one family they linearly mix with all the three up type quarks. In other words the third up quark, the one which is partially composite, is not exactly the top in the basis in which ${(\epsilon_{L,R})}_A\sim\delta_{3A}$: this might induce operators of the form $(\bar{c}_Ru_L)(\bar{c}_Lu_R)$ through the tree level exchange of heavy resonances at their mass scale, which we fix at $1$ TeV. This operator is controlled by
\be
c\simeq{\left(U_d^*\right)}_{32}{\left(V_u\right)}_{31}{\left(V_u^*\right)}_{32}{\left(U_d\right)}_{31}
\ee
and the same bound as before applies here, $c<10^{-8}$, therefore the rotation matrices $V_u$ and $U_d$ cannot be fully generic. A possible way out is to assume that in the discussed basis one of the two top partners, either the right or the left one, does not couple to the quarks of the two other generations; a second possibility is to assume that $V_u$ and $U_d$ have some hierarchical structure which might be related to the CKM matrix or to the hierarchy among families. Both would be consequences of the form of $\lambda_u$ and $\lambda_d$ perhaps explained by physics at the cutoff scale $\Lambda_F$ and we do not discuss them further.

\section{$\SO(9)$ Model with composite $t_R$}\label{top right composto}
\subsection{Structure of the Lagrangian}
As emphasized in \cite{Marzocca:2012zn} in minimal CHM with matter embedded in the fundamental and composite $t_R$ the Higgs mass is predicted to be too light, regardless the presence of SUSY which does not play a role in the argument. In the model we are going to analyze there is an extra source of explicit $\SO(5)$ breaking but we show how it is not sufficient and why also in this case we do not evade the general conclusion.

An explicit realization of a SUSY model where $t_R$ is fully composite is based on a $\SO(9)$ chiral gauge theory with $N_f=9$ flavors \cite{Caracciolo:2012je}, in a way similar to the model of section \ref{top right elementare}. We sketch here the salient features and we refer to the original paper for the details irrelevant for our present discussion.
$\xi^{ia}$ and $\phi^{ia}$, neutral under the gauge group, are the spurion containing $q_L$ and a new exotic field.
When they are decoupled the unbroken global symmetry is $G_f= \SO(5) \times \SU(4)$.
The new field, $\SU(2)_L$ singlet with hypercharge 2 contained in the spurion $\phi$, is necessary for anomaly cancellation. Notice that it is elementary, we cannot take it arising within the strongly interacting gauge theory because the latter is well defined and it is not left unspecified. 
The explicit embeddings in spurions are given in eq.(4.4) of \cite{Caracciolo:2012je}.

The $\SO(9)$ theory becomes strongly coupled at the scale $\Lambda$ and at lower energies it is described by an emergent $\SO(4)_m$ gauge theory.
The elementary fields couple to the $\SO(4)_m$ mesons through the mass mixing $\epsilon_t$ and $\epsilon_\phi$.

The spontaneous SUSY breaking is not enough to give a sizable mass to the SSM sparticles, so we add explicit soft breaking terms to the theory.
We also add SUSY breaking terms in the composite sector, by assuming that they respect the global symmetry $G_f$.

A linear combination of fermions given by $t_L$ and the appropriate components of  $\psi_{M_{im}}$ remains massless and is identified with the SM left-handed top. The right top superfield is contained in the meson $M_{i5}\sim Q_iQ_5$.

\subsection{Numerical Analysis}

The next step is the computation of the effective action for the Higgs field, performed in the unitary gauge. As we mentioned before the Higgs is a NG boson of a spontaneous breaking of a global symmetry: the broken symmetry is not exact and it is explicitly broken by the SM EW gauge group and by the couplings  $\epsilon_t$ and $\epsilon_\phi$. In the matter contribution we further distinguish between colored and non colored exotic fields. We then perform a numerical scan in the parameter space.

The value of $\epsilon_t$ is fixed by the top mass; the mixing $\epsilon_\phi$ of the non colored field is in principle free. $\gamma_g$, $\gamma_m^{(c)}$ and $\gamma_m^{(nc)}$ are equally important and they cancel against each other: the size of these cancellations is a lower bound on the FT. For what concern the coefficient of the quartic term we have $\beta\sim\beta_m\gg\beta_g$.

The Higgs turns out to be too light ($\sim 100$ GeV) unless a sizable source of $\SO(5)$ breaking comes from the non colored sector, as in fact was expected by simple arguments based on general assumptions resumed in \cite{Marzocca:2013fza}. 
Since the Higgs mass square is proportional to the sum $\beta_m^{(c)}+\beta_m^{(nc)}$ in principle raising the non colored contribution controlled by $\epsilon_\phi$ would be sufficient. At the same time large values for $\epsilon_\phi$ are disfavored because generally $\gamma_m^{(nc)}<0$ and it tends to align the Higgs in a EW preserving vacuum. Due to this tension the model as it stands is excluded. We have chosen to report the results because, despite SUSY, the construction is quite minimal and we expect it to be representative for more general examples: it embodies a composite top right model with the addition of an extra massive singlet. The situation can be improved if we introduce more FT: due to the logarithmic dependence on soft masses we would need stops at a scale $O(100)$ TeV, definitely losing the naturalness. We can also introduce more complication in the model or focus on $\SO(5)$ representations different from the fundamental, but we do not continue along this path.

\section{Conclusions}\label{Conclusions}

We have studied SUSY models of composite Higgs, namely we concentrate on numerical results for some models previously introduced. While the specific realization with a fully composite top right does not reproduce the correct Higgs mass value at the chosen reference value of $\xi=0.1$, and not even for reasonably more tuned values, the model with partially composite top quark can accommodate it.

We thus derived bounds on new particles' masses reinterpreting existing searches at LHC. The model is not excluded and interestingly enough some lighter states could be accessible soon at 14 TeV: in particular fermionic partners, especially with exotic hypercharge, would be a smoking gun of composite Higgs model and on top of that our supersymmetric setup would predict the existence of scalar partners in the same range of masses, below the TeV.

Given the kindness of the model we decided to take few steps further and study a possible mechanism to communicate EWSB to all SM quarks and give them masses. It relies on the generation of dimension five operators at a scale chosen at $100-1000$ TeV and on the presence of a slightly more complicated, but more general, vacuum structure. In this way the top mass and other masses are qualitatively different and we account for the observed hierarchy.

The model shares the same tensions coming from flavor constraints as the more conventional framework of the MSSM and no additional troubles are introduced once irrelevant deformations are turned on: we thus expect to employ existing ideas to avoid flavor bounds, as for instance alignment among squark masses.
 
Also we noticed that the presence of top resonances can induce unwanted flavor violations if the UV structure of the model is completely generic: the solution might be related to the origin of the hierarchies in the quark sector and we did not investigate it in detail.

Precision measurements together with already existing searches at colliders steer us to gain useful insights. Future experiments will eventually hint some new physics, either in the form of some direct detection or in some deviations, or will bring us to regions of higher and higher tuning where more radical ideas will be needed. We thus reserve the possibility to better investigate this class of models and make more precise predictions, taking advantage of their self contained validity as effective theories up to scales of the order $100\,-\,1000$ TeV.

\section*{Acknowledgements}
The author is grateful to David Marzocca for collaboration on an early stage of the project; he is also indebted to Marco Serone for discussions and for feedback on the manuscript.
\appendix
\section{Spectrum of $\SO(11)$ Model}
\label{sec:Spectrum}

We present here the particle spectrum of the model in section \ref{top right elementare}, neglecting EWSB effects and the vev eq.(\ref{vev meson}). Throughout the paper we have defined the gauge sector as the one which contributes to the one loop Higgs potential via the SM electroweak gauge couplings, while the matter sector as the one which contributes through the mixings $\epsilon$. This classification reflects also the R-parity assignment for the superfields: it is the same as the one of the corresponding SM superfield with which the field mixes.
\begin{table}[!h]
\begin{center}
\begin{tabular}{c | c c c c | c c c c c}
		& $W, B, G$ & $\rho_m$ & $q_a^n$ & $M_{ab}$ & $q_L$ & $t_R$ & $M_{ia}$ & $q_i^n$ & $M_{ij}$ \\
	 $R_P$& $+$ & $+$ & $+$ & $+$ & $-$ & $-$ & $-$ & $-$ & $+$ \\
\end{tabular}
\caption{$R_P$ assignment of the lowest component of the superfields.}
\label{table:R parity}
\end{center}
\end{table}
\subsection{Gauge Sector}
\begin{table}[h]
\begin{tabular}{|c|c|c|}
\hline
Field & Mass & $(\SU(3)_c, \SU(2)_L, \U(1)_Y)$\\\hline
Real: $ q_{\rho L}$ & $\sqrt{(g_m^2+g_0^2)f^2 - 2 \widetilde{m}_4^2}$ & $(1,3,0)$ \\
Real: $ q_{\rho R}^3$ & $\sqrt{(g_m^2+g_0^{\prime 2})f^2 - 2 \widetilde{m}_4^2}$ & $(1,1,0)$ \\
$ q_{\rho R}^{+}$ & $\sqrt{g_m^2 f^2 - 2 \widetilde{m}_4^2}$ & $(1,1,1)$ + h.c. \\
$ H_d$ & $\sqrt{2 h^2 f^2 - 2 \widetilde{m}_4^2}$ & $(1,2,-\frac{1}{2} )$ + h.c.\\
Reals: $ s_{q}^{1,2}$ & $(\sqrt{2} h f, \;  \sqrt{2 h^2 f^2 - 2 \widetilde{m}_4^2} )$ & $(1,1, 0)$ \\
Reals: $ \phi_{q \pm, 0}^{1,2}$ & $(\sqrt{2} h f , \; \sqrt{2 h^2 f^2 - 2 \widetilde{m}_4^2} )$ & $(1,3,(\pm1,0)) $ \\
$ M_{u,d}$ & $\sqrt{h^2 f^2 + \widetilde{m}_2^2}$ & $(1,2,\pm\frac{1}{2} )$ + h.c.\\
$ s_{M}$ & $\sqrt{2 h^2 f^2 + \widetilde{m}_2^2}$ & $(1,1,0)$ + h.c.\\
$ \phi_{M \pm, 0}$ & $\sqrt{2 h^2 f^2 + \widetilde{m}_2^2}$ & $(1,3,(\pm1,0)) $ + h.c. \\
$ M_{55}$ & $\sqrt{\widetilde{m}_2^2 + \delta \widetilde{m}_{55}^2}$ & $(1,1,0)$ + h.c.\\
\hline
\end{tabular}
\begin{center}\small{(a)}\end{center}
\end{table}
\newpage
\begin{table}[h]
\begin{tabular}{|c|c|c|}
\hline
Field & Mass & $(\SU(3)_c, \SU(2)_L, \U(1)_Y)$ \\
\hline
$\rho^{\pm1}$ & $f g_m $ & $(1,1,\pm 1)$ \\
$\rho^B$ & $f \sqrt{g_m^2 + g_0^{\prime 2}} $ & $(1,1,0)$ \\
$\rho^W$ & $f \sqrt{ g_m^2 + g_0^2}$ & $(1,3,0)$ \\
\hline
\end{tabular}
\begin{center}\small{(b)}\end{center}
\end{table}
\begin{table}[h]
\begin{tabular}{|c|c|c|}
\hline
Field&Mass&$(\SU(3)_c, \SU(2)_L, \U(1)_Y)$\\\hline
$ \tilde \rho_L^a, \tilde q_{\rho L}^a, \tilde w^a$ & $\mathcal{M}_{\tilde w}$ & $(1,3,0)$ \\
$ \tilde \rho_R^3, \tilde q_{\rho R}^3, \tilde b$ & $\mathcal{M}_{\tilde b}$ & $(1,1,0)$ \\
$  \tilde \rho_R^{\pm}, \tilde q_{\rho R}^{\pm}$ & $m_\rho^\pm$ & $(1,1,\pm 1)$ \\
$\tilde h_{u,d}$ & $h f$ & $(1,2,\pm\frac{1}{2})$ \\
$\tilde \phi_{\pm, 0}, \; \tilde s$ & $\sqrt{2} h f$ & $(1,3,(\pm1,0)) + (1,1,0)$ \\
$\psi_{M_{55}}$ & $m_{\psi_{M_{55}}}$ & $(1,1,0)$ \\
$\widetilde{g}$ & $m_{3/2}$ & $(1,1,0)$ \\
\hline
\end{tabular}
\begin{center}\small{(c)}\end{center}
\caption{Spectrum of heavy gauge fields: (a) scalars, (b) vectors, (c) fermions.}
\label{tab1}
\end{table}
{\allowdisplaybreaks\begin{align}
(\mathcal{M}_{\tilde w})^2 &= \left(\begin{array}{ccc}
m_\lambda^2+g_m^2f^2&ig_m m_\lambda f&-g_0^2f\\
-ig_m m_\lambda f&(g_m^2+g_0^2)f^2&ig_0 M_2f\\
-g_0^2f^2&-ig_0 M_2f&M_2^2+2g_0^2f^2
\end{array}\right)\nonumber\\
(\mathcal{M}_{\tilde b})^2&=(\mathcal{M}_{\tilde w})^2\quad\mbox{with}\quad\left\{ g_0\rightarrow g_0',\,\, M_2\rightarrow M_1 \right\}\nonumber\\
(m_{\rho}^\pm)^2 &= f^2 g_m^2+\frac{m_\lambda^2}{2}\pm\frac{m_\lambda}{2}\sqrt{4f^2 g_m^2+m_\lambda^2},\\
\nonumber\\
\mbox{$m_\lambda$ is}&\mbox{ a soft Majorana mass for ${\SO(4)}_m$ gauginos.}\nonumber\\
\mathcal{M}_{\tilde w},\,&\mathcal{M}_{\tilde b}\mbox{ and }m_{\rho}^\pm\mbox{ are Majorana masses.}\nonumber
\end{align}}
\subsection{Matter Sector}
\begin{table}[h]
\begin{tabular}{|c|c|c|}
\hline
Field & Mass & $(\SU(3)_c, \SU(2)_L, \U(1)_Y)$ \\
\hline
$\tilde q_L,q_M,q_q$ & ${\mathcal M}_{q_L}^{1/6}$ & $(3,2,\frac{1}{6})$ \\
$\tilde{Q}^{1/6}_{\pm}$ & $\tilde m_{Q \pm}^{1/6}$ & $(3,2,\frac{1}{6})$ \\
$\tilde{\bar{Q}}^{-1/6}_{\pm}$ & $\tilde m_{\bar{Q} \pm}^{-1/6}$ & $(\bar3, 2, -\frac{1}{6})$ \\
$\tilde{X}^{7/6}_{\pm}$ & $\tilde m_{X \pm}^{7/6}$ & $(3,2,\frac{7}{6})$ \\
$\tilde{\bar{X}}^{-7/6}_{\pm}$ & $\tilde m_{\bar{X} \pm}^{-7/6}$ & $(\bar3, 2, -\frac{7}{6})$ \\
$\tilde t_R, \; \tilde S^{-2/3}$ & $\tilde m_{\tilde S \pm}^{-2/3}$ & $(\bar3, 1, -\frac{2}{3})$ \\
$\tilde{\bar{S}}^{2/3}$ & $\tilde m_{\bar S^{2/3}}$ & $(3, 1, -\frac{2}{3})$ \\
\hline
 \end{tabular}
\begin{center}\small{(a)}\end{center}
\end{table}
\begin{table}[h]
\begin{tabular}{|c|c|c|}
\hline
Field & Mass & $(\SU(3)_c, \SU(2)_L, \U(1)_Y)$ \\
\hline
$Q^{1/6}_{\pm}$ & $m^{1/6}_{Q\pm}$ & $(3,2,\frac{1}{6})$ \\
$X^{7/6}_{\pm}$ & $m^{7/6}_{X\pm}$ & $(3,2,\frac{7}{6})$ \\
$S^{-2/3}$ & $m_S$ & $(\bar 3,1,-\frac{2}{3})$ \\
\hline
 \end{tabular}
\begin{center}\small{(b)}\end{center}
\caption{Spectrum of heavy matter fields: (a) scalars, (b) fermions.}
\label{tab2}
\end{table}
\newpage
{\allowdisplaybreaks\begin{align}
	\left( {\cal M}^{1/6}_{q_L} \right)^2& = \left( \begin{array}{c c c}
	\widetilde{m}^2_{t_L} + \frac{\epsilon_L^2}{2} & - \frac{m_2 \epsilon_L}{2 \sqrt{2}} & \frac{hf}{\sqrt2} \epsilon_L \\
	- \frac{m_2 \epsilon_L}{2 \sqrt{2}} & h^2f^2 + \frac{m_2^2 + \widetilde m_1^2}{4} & - \frac{ hf m_2}{2} \\
	 \frac{hf}{\sqrt2} \epsilon_L & - \frac{hf m_2}{2}  & h^2f^2 + \widetilde m_3^2
	\end{array} \right)\,,\nonumber\\
	(\tilde m_{\bar{Q} \pm}^{-1/6})^2& = \; \frac{1}{8} \left( 8h^2f^2 + m_2^2 + \widetilde m_1^2 + 4 \tilde m_3^2 + 2 \epsilon_L^2 \right) \nonumber\\
	&\quad \pm \frac{1}{8} \left( 16 h^2f^2 m_2^2 + m_2^4 + 2 m_2^2 \widetilde m_1^2 - 8 m_2^2 \widetilde m_3^2  +\right.\nonumber\\
	&  \quad\left. + 4 m_2^2 \epsilon_L^2 + \widetilde m_1^4- 8 \widetilde m_1^2 \widetilde m_3^2 + 4 \widetilde m_1^2 \epsilon_L^2 + 16 \widetilde m_3^4+    	\right.\nonumber\\&\left. - 16 \widetilde m_3^2 \epsilon_L^2 + 	4 \epsilon_L^4 \right)^{1/2}\,,\nonumber\\
	(\tilde m_{X \pm}^{7/6})^2 &= \; (\tilde m_{\bar{X} \pm}^{-7/6})^2 = (\tilde m_{\bar{Q} 1,2}^{-1/6})^2 \text{ with } \epsilon_L = 0\,,\nonumber\\
	(\tilde m_{\bar S^{2/3}})^2 &= \frac{m_2^2 + \widetilde m_1^2 + 2 \epsilon_R^2}{4}\,,\nonumber\\
	(\tilde m_{\tilde{S} \pm}^{-2/3})^2& = \; \frac{1}{8} \left( m_2^2 + \widetilde m_1^2 + 4 \widetilde m_{t_R}^2 + 2 \epsilon_R^2 \right)+ \nonumber\\
	&\quad \pm \frac{1}{8} \left( m_2^4 + 2 m_2^2 (\widetilde m_1^2 - 4 \widetilde m_{t_R}^2 + 2 \epsilon_R^2) \right.+\nonumber\\&\quad\left.+ (\widetilde m_1^2  - 4 \widetilde m_{t_R}^2 - 2 	\epsilon_R^2)^2  \right)^{1/2}\,.\nonumber\\
	(m^{1/6}_{Q\pm})^2 &= \frac{1}{8} \left( m_2^2 + 2 \epsilon_L^2 + 8 h^2 f^2 +\right.\nonumber\\&\left. \pm \sqrt{m_2^4 + 4 m_2^2 \epsilon_L^2 + 4 \epsilon_L^4 + 16 h^2 f^2} \right)\,,\nonumber\\
	(m^{7/6}_{X\pm})^2 &=h^2 f^2 + \frac{m_2}{8} \left( m_2 \pm \sqrt{m_2^2 + 16 h^2 f^2}\right)\,,\nonumber\\
	m_{S} &= \frac{1}{2} \sqrt{m_2^2 + 2 \epsilon_R^2}\,,\nonumber\\
	m_{top} &= \frac{h v}{\sqrt{2} \sqrt{\left( 1 + \frac{2 h^2 v^2}{\xi \epsilon_L^2} \right)\left( 1 + \frac{ m_2^2 }{2 \epsilon_R^2} \right) }}\,.\nonumber\\
\end{align}}


\end{document}